\documentclass[10pt,twocolumn,aps,pra,showpacs]{revtex4}

\usepackage{graphicx}
\usepackage{amssymb}
\usepackage{amsmath}

\usepackage{epstopdf}

\begin{document}

\title{Energy distribution and cooling of a single  atom \\
in an optical tweezer}
\author{C.~Tuchendler, A.~M.~Lance\footnote{Electronic address: andrew.lance@institutoptique.fr},  A.~Browaeys, Y.~R.~P.~Sortais and P.~Grangier} 
\affiliation{Laboratoire
Charles Fabry de l'Institut d'Optique, CNRS, Univ.Paris-Sud, Campus
Polytechnique, RD 128, 91127 Palaiseau Cedex, France}

\date{\today}

\begin{abstract}
We investigate experimentally the energy distribution of a single $^{87}{\rm Rb}$ atom trapped in a strongly 
focused dipole trap under various cooling regimes. Using two different 
methods to measure the mean energy of the atom, we 
show that  the energy distribution of the radiatively cooled atom is close to thermal. 
We then demonstrate how to reduce  the energy of the single atom, 
first by adiabatic cooling, and then by truncating the Boltzmann distribution of the single atom. 
This provides a non-deterministic way to prepare atoms at low microKelvin temperatures, 
close to the ground state of the trapping potential. 
\end{abstract}

\maketitle

\section{Introduction} \label{sec:Intro}

In the last thirty years, the manipulation of single quantum objects has been the 
subject of considerable attention. After the  pioneering experiments of  H. Dehmelt with a single electron in a Penning trap~\cite{Dehmelt90}, a single ion was trapped in a Paul 
trap~\cite{Neuhauser1980}, which led to the observation of quantum 
jumps~\cite{Nagourney86,Bergquist86}. Several techniques were then developed to cool
the single trapped particle, 
such as optical sideband cooling, which was experimentally demonstrated in 1989~\cite{Diedrich89}.
More recently, the interest in  the manipulation of single quantum objects has grown 
with the recognition that they can carry quantum information, and that they therefore can be a 
resource for  quantum information processing~\cite{Nielsen}. In this context, single trapped laser cooled ions have accumulated an impressive 
amount of results in recent years (see e.g.~\cite{QFFT}).

The trapping of single neutral atoms, on the other hand, is more recent, owing in 
part to their weaker interaction with the electromagnetic field. So far, the trapping of 
neutral atoms
has been realized in tight magneto-optical traps~\cite{kimble,ertmer}  or
optical dipole traps~\cite{Frese00, Schlosser2001, Weber2006,Yavuz06}.
As the trapping potentials  are typically less than a milliKelvin deep, laser cooling 
is the usual starting point of the manipulation of single atoms. As is the case for 
ions, it is tempting to further reduce the energy of the trapped atom, 
to ultimately reach the ground vibrational state of the trapping potential. 
In the framework of quantum computing, for example, the entanglement of 
two atoms via controlled collisions usually requires ground state cooling
(see e.g.~\cite{Mandel03b}).

The knowledge and control 
of the energy of a single atom, trapped in a sub-micron optical dipole trap, is  important for many purposes. For instance, in a recent experiment~\cite{Beugnon06}, we have used two single rubidium 
atoms, trapped in two optical tweezers separated by a few microns, as single-photon 
sources. When the two photons (each photon coming from one atom) are superimposed on a  beamsplitter, we observe a 
``coalescence" effect~: the two (indistinguishable) photons copropagate from the same output of the beamsplitter, due to a two-photon interference effect. We have shown that the ultimate visibility of this effect  is limited by the spread in frequencies of the emitted photons,  which in turn  is due to the spread in energy of the emitting trapped atoms.
In another experiment~\cite{Jones07}, using Raman 
transitions we have prepared a single atom 
in a superposition of two internal states of the hyperfine manifold, and we have studied the dephasing of this quantum bit. In this case, due to the fact that successive atoms have different energies, 
each realization of the qubit evolves differently with respect to a local oscillator. 
This results in a  loss of coherence, that we were nevertheless able to compensate 
for by applying rephasing ``spin-echo'' techniques. For this application, a lower 
energy spread of the trapped atom would also increase the dephasing time, thus avoiding the extra ``spin-echo''  sequence.
   
In the present paper, we analyze in more detail the methods used 
in these previous works to determine the mean energy of a single atom
in the optical tweezer \cite{Beugnon06, Jones07, Sortais07}. Similar or complementary techniques have  been reported by other groups~\cite{Alt03, Weber2006}.  
We then exploit these temperature measurements  to 
characterize several cooling methods to further decrease the temperature of the atoms, 
and we show how to approach the ground state level
in the two radial directions of the trap, which exhibit the strongest confinement.

The paper is organized as follows. We describe in section~\ref{sec:ExptSetup} the 
experimental setup. In section \ref{sec:RRMethod}, we present a  
``release and recapture'' technique used to measure the temperature 
of the atoms. In section \ref{sec:RadCooling} we develop a cooling sequence 
used to laser-cool the atom in the tweezer. In section~\ref{sec:AdiabaticMethod}, 
we perform the spectroscopy of the energy distribution of the atom in the tweezer 
and find that this distribution is very close to a Boltzmann distribution. 
Section~\ref{sec:TrucBoltz} describes how  the mean energy
of the single atom  can be decreased by truncating this Boltzmann distribution. This provides a  
non-deterministic method to prepare a single atom close to the ground state 
of the trapping potential. In section \ref{sec:AdiabCool} we 
reduce further the temperature of the atoms by adiabatically lowering the potential trap. 
Finally,  we conclude by discussing some implications of these results.

\section{Experimental Setup}  \label{sec:ExptSetup}

\begin{figure}
\begin{center}
\includegraphics[width=8.5cm]{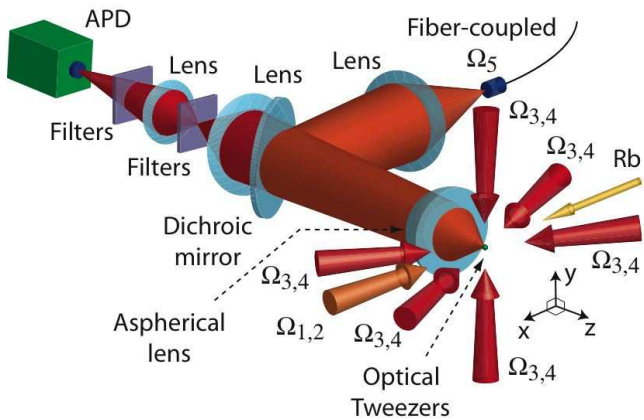}
\caption{(Color online) Schematic of the experimental setup. A beam of $^{87}{\rm Rb}$ atoms, 
originating from an oven, is slowed using a Zeeman coil
(not shown) and counter propagating slowing  $(\Omega_1)$ 
and repumping $(\Omega_2)$ lasers at $\sim780$~nm. An optical molasses is formed at the intersection of 
six counter propagating lasers $(\Omega_3)$ and repumpers $(\Omega_4)$ 
operating  also at $\sim780$~nm. An aspherical lens is used to tightly focus a 
$\sim850$~nm laser $(\Omega_5)$ in the region of the optical molasses.  
The fluorescence of the atom, due to the molasses and Zeeman lasers, is collected 
using the same aspherical lens. A dichroic mirror together with other spatial and spectral filters are 
used to filter the fluorescence signal, which is detected using single-photon counting avalanche photodiode (APD). 
}\label{fig:ExptSetup}
\end{center}
\end{figure}

Figure~\ref{fig:ExptSetup} shows a schematic of the experimental setup. 
A beam of $^{87}{\rm Rb}$ atoms is slowed  using a Zeeman slowing technique~\cite{Phillips82}. 
The counter-propagating laser beam at  $\sim780~{\rm nm}$  (denoted $\Omega_1$) 
is $18.7~\Gamma$ red detuned from the $F=2\to F^{\prime}=3$ cycling transition, 
and combined with a repumping laser $(\Omega_2)$ that is $3.7~\Gamma$ red detuned 
from the $F=1\to F^{\prime}=2$ repumping transition. 
Here, $\Gamma \approx 2\pi\times 6\times10^{6}$~rad/s
is the linewidth of the $D_{2}$ transition. 

We form an optical molasses 
at the intersection of six counter-propagating cooling lasers 
that are $4.2~\Gamma$ red detuned 
from the $F=2\to F^{\prime}=3$ transition  (denoted $\Omega_3$), together with 
six repumping lasers $(\Omega_4)$ resonantly tuned to the $F=1\to F^{\prime}=2$ 
transition. This optical molasses is used as a reservoir of cold atoms from 
which the dipole trap is loaded.  

We produce the optical dipole trap by focusing a $\sim850$~nm laser in the center 
of the molasses, using  a large numerical aperture (${\rm NA}=0.5$) 
aspherical lens~\cite{Sortais07}. The optical dipole trap has a measured 
optical waist of $w = 1.03\pm{0.01}~\mu{\rm m}$. A comprehensive characterization 
of the optical system and of the dipole trap itself is presented in detail 
elsewhere~\cite{Sortais07}. Assuming a Gaussian beam profile in the 
transverse direction for the dipole trap and a Lorentzian profile along the 
propagation axis, we calculate the radial and axial frequencies and find
$\nu_{\perp}\sim 160 $ kHz and $\nu_{\parallel}\sim 30$ kHz respectively 
for 10 mW of laser power. The potential trap depth, proportional to the laser 
power, is $U\sim2.8$ mK for a 10 mW of laser power. 

This optical tweezer allows us to trap single ${\rm Rb}^{87}$ atoms via a 
collisional blockade mechanism, which prevents two or more atoms from being trapped 
simultaneously due to inelastic collisions \cite{Schlosser2001, Schlosser2002,Sortais07}. 
The $\sim780$~nm fluorescence of the trapped single atoms is detected using 
the same aspherical lens that is used to create the dipole trap.  
This fluorescence is filtered using a dichroic mirror, 
together with other spatial and spectral filters, and then 
detected using a single 
photon counter (Avalanche Photodiode), with an overall collection 
and detection efficiency that is estimated to be $\sim0.5\%$.  
From the output signal of the single-photon counter we distinguish between the presence 
(or absence) of an atom in the tweezer, within a 10~ms acquisition period, with a 
confidence better than 99\%. Each experimental sequence that 
will be presented in this paper is triggered on a detected single atom in the dipole trap. 

Our goal in the following sections will be to investigate the energy distribution 
and temperature of the atoms trapped in the optical tweezer, under various
cooling conditions.

\section{Release and Recapture Technique}  \label{sec:RRMethod}

\begin{figure}
\begin{center}
\includegraphics[width=8.5cm]{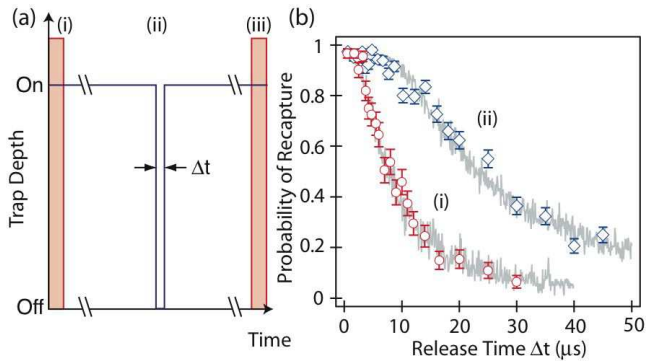}
\caption{
(Color online) Measurement of the temperature of the atoms 
using the ``release and recapture'' method. (a) Schematic of the experimental 
sequence (see text). (b) Experimental results showing the probability of 
recapturing  the atom, $P_{\rm rr}(\Delta{t})$, as a function of release time 
$\Delta{t}$. (i) (Circles) Temperature of the atoms just after the loading 
in the dipole trap. Each data point is the accumulation of 100 sequences.  
Superimposed on this data is  a fit by the Monte-Carlo simulation of the 
``release and recapture'' method, which is the average of 200 trajectories for 
each release time.
The temperature is $168\pm6~\mu$K for a trap 
depth of $\sim2.8$~mK. (ii) (Diamonds) For comparison we have also plotted 
the results for the temperature measurement using the same method, but 
after the atom has been laser-cooled in the tweezer. Each data point is the 
accumulation of 200 sequences. The best-fit temperature is $31\pm1~\mu$K 
for a trap depth of $\sim2.5$~mK.}
\label{fig:TOF_hot_cold_data}
\end{center}
\end{figure}

In this section we will describe a ``release and recapture''  method that we use 
to determine the temperature of the atoms in the dipole trap \cite{Beugnon06, Jones07}. 
We emphasize that what we mean by the temperature of the atoms is a temperature extracted by averaging the energy over many realizations of the same experiment with a single atom.
The premise of this  method is that  information can be obtained about the energy distribution
of a single atom by switching off the dipole trap for a variable time $\Delta{t}$ (as illustrated in 
Fig.~\ref{fig:TOF_hot_cold_data}(a)) and then determining the probability of 
recapturing the atom, which is denoted as $P_{\rm rr}(\Delta{t})$. On average, an 
atom with a high energy  is more likely to escape during the release time 
$\Delta{t}$, compared to an atom with a low energy, as is well-known 
in the context of laser cooling of atomic samples~\cite{Chu85,Lett88}. 
In the single atom regime used here, this method is  particularly robust, since on each repetition 
of the experimental sequence the result is binary~: either the atom has escaped 
the trapping region,  or it is recaptured. This binarization of the measurement outcome
makes it immune to various noise sources, in particular to fluctuations in the fluorescence 
level of the atom, which is used to decide on the presence of the atom.
 
Figure~\ref{fig:TOF_hot_cold_data}(a) illustrates a schematic of 
the ``release and recapture" sequence, which goes as follows.
(i) After the atom is trapped, the cooling and loading lasers ($\Omega_{\rm 1}$, 
$\Omega_{\rm 2}$, $\Omega_{\rm 3}$ and $\Omega_{\rm 4}$) are switched off. 
(ii) The dipole trap laser ($\Omega_{\rm 5}$) is then turned off for a time $\Delta{t}$. 
(iii) Finally, we perform a fluorescence detection by switching all the cooling and loading 
lasers back on, thus determining whether the atom has escaped or not.  
Figure~\ref{fig:TOF_hot_cold_data} shows typical experimental results for 
the ``release and recapture'' technique, where the probability of recapturing 
the atom is plotted as a function of the release time $\Delta{t}$ (more details about this 
experiment will be presented in the following section). 

In order to extract the 
temperature from these measurements, we perform Monte-Carlo simulations of 
the trajectories of the single atoms,  to determine the probability 
that the atoms are recaptured 
after a release time $\Delta{t}$. 
This simulation requires the knowledge of the energy distribution
of the atom in the trap. We assume that we are in the harmonic approximation 
of the trapping potential, and that the position-velocity distribution of the atom
follows a thermal Maxwell-Boltzmann law. The 
standard deviations of the position of the atom in the axial and radial directions 
of the trapping potential are 
$\Delta{x}_{\parallel}=\sqrt{k_{\rm B}T/m \omega_{\parallel}^2}$ 
and $\Delta{x}_{\perp}=\sqrt{k_{\rm B}T/m \omega_{\perp}^2}$ 
respectively, while the standard deviation of the velocity is   
$\Delta{v}=\sqrt{k_{\rm B}T/m}$, where $T$ is the temperature.
The parameters of the optical trapping potential have been determined
previously~\cite{Sortais07}, and  take into account the tilt due to gravity, which 
creates a local maximum in the potential along the gravity axis. The effective trap depth is defined as the height of this potential barrier with 
respect to the minimum of the potential. 

The simulation goes as follows: using the Maxwell-Boltzmann distribution
we randomly generate the three-dimensional 
position-velocity vector 
$(x_{\rm i}, y_{\rm i}, z_{\rm i}, v_{\rm x,i}, v_{\rm y,i},v_{\rm z,i})$ 
of the atom in the potential for a given temperature. We then calculate the trajectory 
of each atom during the release time $\Delta{t}$ in a single time-step 
calculation. The position of the atom after this release time is 
$(x_{\rm i}+v_{\rm x,i}\Delta{t},y_{\rm i}+v_{\rm y,i}\Delta{t}-g\Delta{t}^2/2,z_{\rm i}+v_{\rm z,i}\Delta{t})$, 
where the potential gradient due to gravity along the y-axis has been  included.
We finally determine the energy of each atom when the trap is turned back 
on. If the total energy of the atom after the time-of-flight is smaller than the 
effective trap depth, we consider that the atom is recaptured, otherwise it is 
considered to have escaped. 

By simulating many trajectories of the single atoms for each release time 
$\Delta{t}$, we can numerically determine the probability of recapturing 
the atom as a function of this release time. These simulations are repeated
for a range of different temperatures. We calculate the weighted least 
square value ($\chi^2$) between the experimental data and the simulated 
results for each temperature. The chi-square value is defined 
as $\chi^2=\sum_{\rm i}\{[f_{\rm i}-P_{\rm rr}(\Delta{t}_{\rm i})]^2/\sigma^2_{\rm i}\}$  
where $P_{\rm rr}(\Delta{t})$ is the experimental data that has 
an uncertainty $\sigma$ and $f$ is the value predicted by the model. 
The best-fit temperature corresponds to 
the Boltzmann distribution that minimizes this least square value. 

We performed the ``release and recapture'' technique to determine the 
temperature of the atoms just after they were loaded in the trap. 
Figure~\ref{fig:TOF_hot_cold_data}(b) shows the Monte-Carlo simulation that 
best fits the corresponding experimental data. In this example, the best fit 
temperature is $T=168\pm6~\mu{\rm K}$ in a trap depth of $U\sim2.8$ mK. 
The ratio between the trap depth and the temperature of the atoms is 15, 
indicating that we are approximately in the harmonic regime of the trapping potential. 

Due to experimental miscounts and losses, the probability of recapturing an atom for zero 
release time was less than unity, being instead typically 0.95. We scaled  the Monte-Carlo 
simulation by this factor to achieve a better fit to the experimental
results. The uncertainty on the temperature that minimized the chi-square value 
is defined as $\sigma_{\rm T}=\sqrt{2(\partial^2\chi^2/\partial{T}^2)^{-1}}$, 
assuming that the $\chi^2$ function varies harmonically around 
this minimum point~\cite{Bevington}. Furthermore, we have included 
in the error budget for the temperature, the statistical uncertainty in fitting a parabola 
function to this chi-square minimum.  

We will use this ``release and recapture" technique throughout the rest of the paper 
as a diagnostic to measure the temperature of the atoms in the optical tweezer. 
 
\section{Radiative cooling of a single atom}  \label{sec:RadCooling}

In this Section, we discuss  the laser cooling of a single atom after it has 
been captured in the dipole trap. 
For that purpose,  the 
cooling sequence was optimized by controlling the molasses lasers parameters
(cooling time, frequency detuning, intensity), as well as the trapping laser intensity.  
This was done  by maximizing the recapture probability 
of the single atom after a ``release and recapture" experiment with $\Delta{t}=10~\mu$s.

\begin{figure}
\begin{center}
\includegraphics[width=8.5cm]{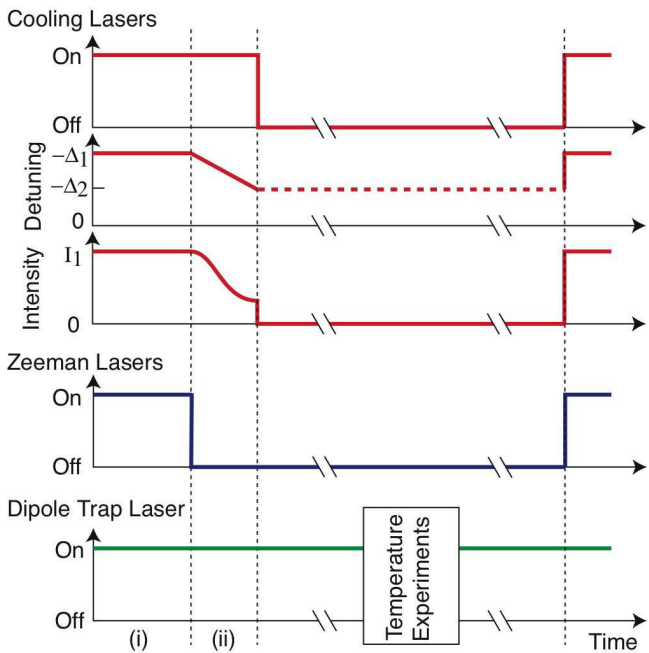}
\caption{(Color online) Schematic of the temporal sequence for the laser 
cooling sequence. The details are given in the body of the text. 
The box labeled {\it Temperature Experiments} shows the point in the 
sequence where we can manipulate the intensity of the dipole trap laser 
in order to measure the temperature of the atoms. 
(See Sections  
\ref{sec:AdiabaticMethod}, \ref{sec:TrucBoltz} 
and \ref{sec:AdiabCool}).}\label{fig:CoolingSequence}
\end{center}
\end{figure}

Figure~\ref{fig:CoolingSequence} shows the optimized laser cooling sequence.
Initially, all lasers are on and the molasses cooling 
lasers are detuned by $\Delta_{1}=-4.5\Gamma$ to maximize 
the loading rate of the molasses and optical dipole trap. 
(i) The sequence is triggered on the detection of an atom, where the 
dipole trap laser power is $\sim10$~mW. (ii) The Zeeman laser and 
the Zeeman repumper laser are switched off. At the same time, the 
cooling lasers are linearly detuned from $\Delta_{1}=-4.5\Gamma$ 
to $\Delta_{2}=-2\Gamma$ in $\sim1.2$ ms. During this time, the 
intensities of the molasses lasers are reduced from 
$I_{1}\sim2$~mW/beam, to approximately one third of their initial intensity. 
After this cooling period the cooling and repumping lasers are switched off.

Using this optimized laser-cooling sequence we were able to significantly 
reduce the temperature of the atoms. Figure~\ref{fig:TOF_hot_cold_data}(b) 
shows the  ``release and recapture'' experimental results after the 
laser-cooling, together with the best-fit simulation results. 
This  corresponds to a temperature of  $31\pm1~\mu$K 
for a trap depth of $\sim2.5$~mK, which yields the ratio  $U/T\sim81$. 
This represents more than a factor 5 reduction in the temperature with respect to 
the temperature of the atoms directly loaded from the optical molasses. 

We note that the optimized cooling sequence is achieved when the 
laser detuning is brought closer to resonance with respect to the atomic transition. 
This fact may seem inconsistent with the sub-Doppler theory of laser cooling 
(see e.g. \cite{Drewsen94}), which predicts that the temperature is lower for larger 
detuning. However one has to take into account that when the atom is trapped in 
the tweezer, the resonance frequency  is  light shifted by about 10$\Gamma$ with 
respect to the free space case. Therefore the detuning experienced by the trapped 
atom varies only from $-14.5\Gamma$ to $-12\Gamma$. The extra cooling effect is then 
attributed to  the decrease (by a factor 3) in the  intensity of the laser beams, rather 
than to the frequency change, now in reasonable agreement with sub-Doppler laser cooling. 
 
This optimized cooling technique is the starting point for each experiment 
presented throughout the rest of the paper. We will now turn our attention to 
validating the assumption about the thermal distribution. In addition,
we would like to cross-check the temperature results using an alternative method. 

\section{Energy distribution of a single atom in an optical tweezer} \label{sec:AdiabaticMethod}

In this section we determine the energy distribution of the single atom in the dipole trap. 
Provided that this distribution is indeed thermal,  the 
corresponding temperature can be evaluated. To achieve this, we apply a type of spectroscopy 
method proposed and experimentally demonstrated by Alt {\it et al.} in ref.~\cite{Alt03}. 

\begin{figure}
\begin{center}
\includegraphics[width=8.5cm]{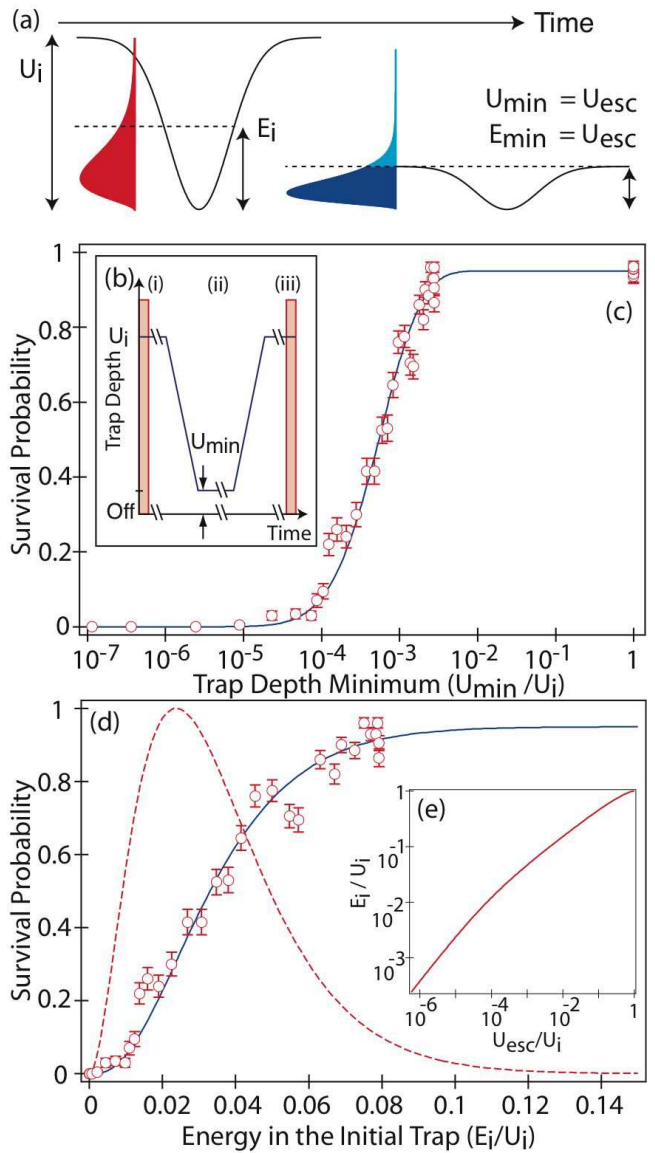}
\caption{(Color online) Spectroscopy of the energy distribution of a 
single atom using the adiabatic lowering method. (a) Schematic showing
that due to the adiabatic lowering of the dipole trap, an atom with initial 
energy $E_{\rm i}$ in a trap depth $U_{\rm i}$ eventually has an energy 
equal to the trap depth, i.e. $E_{\rm min}=U_{\rm min}=U_{\rm esc}$. (b) Schematic 
of the experimental sequence (see text). (c) Experimental results of the survival 
probability for various minimum trap depths $U_{\rm min}$. Each data point is the 
accumulation  of 200 sequences, with error bars due to the binomial statistics.  
Solid line is a theoretical fit to the data.
(d) Reconstruction of the Boltzmann distribution in the  initial trap of depth
$U_{\rm i}$. Plot showing the percentage of atoms with an energy less
than $E_{\rm i}$ (in the initial trap of depth $U_{\rm i}$) versus the 
normalized energy $E_{\rm i}/U_{\rm i}$. Solid line is a theoretical fit 
to the data (see text). Dashed line 
represents the corresponding Boltzmann distribution (e) Numerical results for 
solving the constant action equation 
$S(E_{\rm i},U_{\rm i})=S(U_{\rm esc},U_{\rm esc})$. 
}\label{fig:2ndMethod}
\end{center}
\end{figure}

This method involves the adiabatic lowering of the trap depth to a point where the 
atom can potentially escape. As shown in Fig.~\ref{fig:2ndMethod}(a), by lowering 
the trap depth, an atom with initial energy $E_{\rm i}$ in a trap of depth 
$U_{\rm i}$ will eventually have an energy $E_{\rm esc}$ that is equal to the
final trap depth $U_{\rm esc}$ and hence the atom will escape the trap. Thus, 
measuring the depth at which the atom escapes the trap yields information 
about the initial energy of the atom.

Figure~\ref{fig:2ndMethod}(b) shows a schematic of the experimental sequence 
for this method, which goes as follows: (i) The single atom is trapped and
laser cooled. (ii) The dipole trap depth is then adiabatically lowered from 
$U_{\rm i}\sim 2.8$ mK to $U_{\rm min}$. The lowering time is approximately 
$\sim2.5~{\rm ms}$.
The trap depth is held 
constant for a duration of $\sim20~{\rm ms}$, to allow an atom with an 
energy greater than the trap depth sufficient time 
to escape~\footnote{We assume that an atom with an 
energy greater than the trap depth has sufficient time to exit the trap
provided the following condition is fulfilled: $1/\tau\ll\nu_{\parallel}$, 
where $\tau$ is the time the atom is in the trap and $\nu_{\parallel}$ is the 
trap frequency in the axial direction. This condition is satisfied down to a final
trap depth $U_{\rm f}/U_{\rm i}\sim10^{-4}$, where 
$\tau\times\nu_{\parallel}\sim10\gg1$.}.
(iii) After this waiting time, the trap depth is adiabatically raised to 
the initial trap depth $U_{\rm i}$ in $\sim2.5~{\rm ms}$. Finally, the laser beams 
are switched back on to determine if the atom has escaped or not. We leave 
to section~\ref{sec:AdiabCool} the discussion about the adiabaticity of this 
process. Here we assume that the hypothesis for the adiabatic change 
of the potential is fulfilled.
Figure~\ref{fig:2ndMethod} (c) shows the experimental probability of recapture 
for various minimum trap depths $U_{\rm min}$ normalized to the initial trap depth $U_{\rm i}$
\footnote{For the effective minimum trap depth, $U_{\rm min}$, taking gravity into 
account turns out to be important for trap depths less than 
$\sim10~\mu$K because the gravity potential 
acts to tilt the trapping potential, thus 
lowering the potential barrier along the gravity axis.}. 

It is now necessary to map the minimum trap depth $U_{\rm min}$ back to the 
corresponding energy in the initial trap $E_{\rm i}$. To perform this mapping we 
calculate the one-dimensional action that can be 
expressed as the integral $S(E,U)=\int^{x_{\rm max}}_{0}\sqrt{2m[E-U(x)]}dx$, 
where $E$ is the energy, $U(x)$ is the potential and $x_{\rm max}$ is the 
position where the atom has zero kinetic energy. In the adiabatic limit, the 
action of the atom in the trap is conserved as the trap depth is adiabatically 
lowered~\cite{Landau}. We solve the constant action equation 
$S(E_{\rm i},U_{\rm i})=S(U_{\rm esc},U_{\rm esc})$ in the radial 
direction of our potential trap that includes the gravity potential 
gradient.
The numerical results of the $U_{\rm esc}/U_{\rm i} \to E_{\rm i}/U_{\rm i}$ 
mapping is plotted in Figure~\ref{fig:2ndMethod}~(e).
We subsequently apply the mapping $U_{\rm esc}/U_{\rm i} \to E_{\rm i}/U_{\rm i}$ 
to our experimental data. Figure~\ref{fig:2ndMethod}~(d) shows the percentage 
of atoms that have an energy less than $E_{\rm i}$ 
in the initial trap of depth $U_{\rm i}$ plotted as a function of the normalized 
energy $E_{\rm i}/U_{\rm i}$. 

We now wish to compare the measured energy distribution to a thermal one. 
In the harmonic trap limit (i.e. when the ratio of the trap depth to the temperature 
of the atoms satisfies $U_{\rm i}/ k_{\rm B}T\gg1$), we calculate the Boltzmann energy distribution for a temperature $T$. We assume that the normalized energy distribution of the atom is 
\begin{equation}\label{eqn:ProbBoltz}
f_{\rm th}(E)=\frac{1}{2(k_{\rm B} T)^3}E^2 e^{-E/{k_{\rm B} T}}
\end{equation}
where ${E}^2$ corresponds to the three dimensional density of states. The probability that an atom has an energy less than $E$ is 
defined by $P_{\rm surv}(E)=\int^{E}_{0} f_{\rm th}(E^{\prime})~dE^{\prime}$ and is given by
\begin{equation} \label{eqn:ProbSur}
P_{\rm surv}(E)=1-\Big[1+\eta+\frac{1}{2}\eta^2 \Big] e^{-\eta}
\end{equation}
where $\eta=E/{k_{\rm B} T}$. Equation~(\ref{eqn:ProbSur}) can be interpreted 
as the survival probability of the atom remaining in the dipole trap after the truncation of the Boltzmann 
distribution.  

Figure~\ref{fig:2ndMethod}(d) shows a fit of the integrated Boltzmann distribution 
$P(E_{\rm i}/U_{\rm i})$ to the experimental data using Eq.~(\ref{eqn:ProbSur}). To achieve a better fit to the 
data, the maximum survival probability is rescaled to $0.95$, to account for non-ideal 
recapture probability of our experiment when $U_{\rm min} =U_{\rm i}$.
We find that the measured energy distribution is well fitted by a thermal distribution. This 
fact shows that over many realizations of the experiments, the energy of successive 
atoms after laser cooling is well described by a  Boltzmann distribution, 
in agreement with  several previous theoretical and experimental studies ~\cite{Lett89}.

The Boltzmann distribution in Figure~\ref{fig:2ndMethod}(d) has a 
corresponding temperature of $T~=~33\pm~2~\mu{\rm K}$ for a trap 
depth $U_{\rm i} \sim 2.8$ mK, where the error in the temperature 
corresponds the statistical uncertainty in the fit. 
This temperature is in good agreement with the ``release and recapture" temperature 
results presented in the previous section.

We point out that the adiabatic lowering of the dipole trap depth is a powerful 
tool in manipulating the energy distribution of the atom. For example, at very low 
trap depths the more energetic atoms will escape with a higher probability. This 
adiabatic lowering indeed acts as a filter such that only the coolest atoms will remain 
in the trap, as we will now show. 

\section{Reduction of mean energy by truncating the Boltzmann distribution} \label{sec:TrucBoltz}

\begin{figure}
\begin{center}
\includegraphics[width=8.5cm]{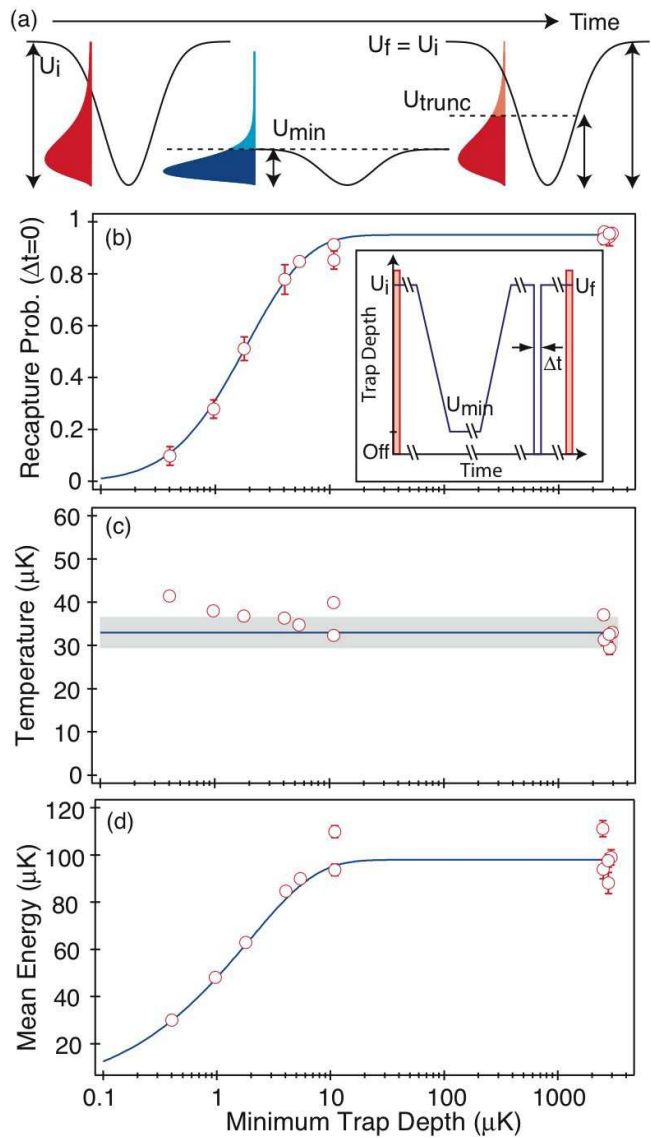}
\caption{
(Color online) Investigation of a truncated Boltzmann distribution using the 
``release and recapture'' method. (a) Schematic of the experimental sequence. 
(b) Probability of recapture for 
zero release time $P_{\rm rr}(\Delta{t}=0)$ plotted as a function of the minimum 
trap depth $U_{\rm min}$. $P_{\rm rr}(\Delta{t}=0)$ is defined as the mean 
of the first four recapture probabilities with the smallest release times of a given 
data set (e.g. see Fig.~\ref{fig:TOF_hot_cold_data}(b)), where
the error bars are due to normal statistics.  
Solid line is a theoretical fit 
to the data.  
Inset shows a schematic of the temporal experiment sequence. 
(c) The temperature $T$ plotted as a function of $U_{\rm min}$. 
Solid line corresponds to a constant temperature of $33~\mu$K.
Grey bar corresponds to the standard deviation of the 
spread in the measured temperatures for the minimum trap depth 
$U_{\rm min}\sim2.8$~mK.
(d) The calculated mean energy $\langle E\rangle$ of the 
truncated Boltzmann distribution shown as a function of $U_{\rm min}$. 
The solid line 
is a theoretical prediction of the mean energy of a truncated Boltzmann distribution
with a temperature of $33~\mu$K.
}\label{fig:PostSelectResults}
\end{center}
\end{figure}

We have  shown in Sections~\ref{sec:RadCooling} and \ref{sec:AdiabaticMethod} 
that we can cool the atoms down to a temperature of approximately 
$T=33~\mu{\rm K}$ using laser cooling techniques. It is possible to further reduce
the mean energy of the trapped atom by truncating the Boltzmann distribution at 
an energy $U_{\rm trunc}$. Here, we perform this truncation by adiabatically 
lowering the trapping potential, to a point where atoms with a higher energy are 
likely to escape, while atoms with a lower energy remain trapped.
This filtering via the lowering of the trap potential appears to be similar to evaporative 
cooling, but a crucial difference is that there is no collision-induced rethermalization, 
since the experiment is done at the single atom level~\cite{Luiten96}.

Figure~\ref{fig:PostSelectResults}(a) shows a schematic of the experimental 
sequence that we use to investigate the energy distribution of the truncated 
Boltzmann distribution. This experimental sequence is identical to the one 
described in section~\ref{sec:AdiabaticMethod}, but we now perform a ``release
and recapture" sequence directly after it. The entire temporal sequence is shown in 
the inset of figure~\ref{fig:PostSelectResults}(b). The additional ``release and recapture" 
sequence allows us to determine the temperature of the atoms after the truncation 
process and after the trap depth has been returned to the initial depth.

To determine the truncation energy $U_{\rm trunc}$ in the trap of 
potential depth $U_{\rm f}=U_{\rm i}$, we first calculate the truncation energy in the shallow 
trap of potential depth $U_{\rm min}$ (using the measured minimum optical dipole trap power). 
We then use the numerical results that map $U_{\rm min}\to E_{\rm i}$ (see Fig.~\ref{fig:2ndMethod}(e)) 
to determine the  truncation energy $U_{\rm trunc}$ in the trap 
of potential depth $U_{\rm f}$.

In the case of a truncated Boltzmann distribution, it is not sufficient 
to assume that the position-velocity vector 
of the atom follows a Gaussian Maxwell Boltzmann distribution, as we did for 
the  ``release and recapture" experiments in Sections~\ref{sec:RRMethod} 
and \ref{sec:RadCooling}. To account for this effect, in the following simulations 
we consider that the energies of the atoms are drawn from a truncated Boltzmann 
distribution corresponding to a temperature $T$ and a truncation energy 
$U_{\rm trunc}$. The Boltzmann 
distribution is discretized into $N$ bins with energies $E_{\rm j}$ equally spaced by 
$\Delta{U}=U_{\rm trunc}/N$, for $\Delta{U}/2\leq E_{\rm j}\leq(U_{\rm trunc}-\Delta{U}/2)$. 
For each distribution, the number of simulations with atoms with an energy 
$E_{\rm j}$ is weighted according to this discretized Boltzmann distribution. 
For an atom with energy $E_{j}$, we randomly distribute the energies 
among the three directions given by the cartesian axes. 
The period of motion of the atom is then randomly chosen
for each of the three cartesian axes (ensuring that the total energy of the atom remains $E_{j}$). 
Finally, the trajectories of the atoms  are simulated during the time of flight as described 
in Section~\ref{sec:RRMethod}. We typically use several hundred atom 
trajectories for each release time $\Delta{t}$, and we discretize the Boltzmann 
distribution into $10$ bins. When the truncation energy is very 
large with respect to the temperature (i.e. the distribution is approximately 
not truncated), this new model yields the same results as the previous release 
and recapture model. 

The experimental results are shown in Figure~\ref{fig:PostSelectResults}. 
We repeated the experiment several times for different minimum trap depths
$U_{\rm min}$. The data were recorded over a span of several days, where all 
other parameters were kept approximately constant. 
For each minimum trap depth we measured the recapture 
probability, $P_{\rm rr}(\Delta{t}=0)$, 
the temperature, $T$, of the truncated Boltzmann distribution
and we calculated the associated mean energy, $\langle E\rangle$, of this distribution.
The small day to day variations in the experimental 
parameters result in a small spread of the temperatures for a given value of 
$U_{\rm min}$. Nevertheless, Figure~\ref{fig:PostSelectResults}(c) shows 
that the temperature of the Boltzmann distribution remains approximately 
constant down to very small trap depths, even when the 
distribution becomes truncated.   

In contrast, the mean energy is reduced markedly as the 
distribution becomes more truncated, as we will now describe.  
The mean energy of a truncated 
Boltzmann distribution in a three dimensional harmonic potential 
can be expressed as:
\begin{eqnarray} \nonumber
\langle E\rangle&=&\frac{\int^{U_{\rm trunc}}_{0} E^3\ e^{-E/k_{\rm B}T}~dE}
{\int^{U_{\rm trunc}}_{0} E^2\ e^{-E/k_{\rm B}T}~dE} \\
&=& 3k_{\rm B} T\Bigg[\ 
\frac{1-\Big(1+\eta+\frac{1}{2}\eta^2 +\frac{1}{6}\eta^3 \Big)e^{-\eta} }
{1-\Big(1+\eta+\frac{1}{2}\eta^2 \Big)e^{-\eta} }\Bigg] \label{eqn:MeanEnergy} 
\end{eqnarray}
where $\eta=U_{\rm trunc}/{k_{\rm B} T}$. In the limit where 
the Boltzmann distribution is approximately non-truncated, i.e. $\eta\gg1$, 
the mean energy is given by $\langle E\rangle=3k_{\rm B} T$.
However, as shown in Eq.~(\ref{eqn:MeanEnergy}) the mean energy is reduced 
significantly as the truncation is increased, i.e.~$\eta\to 0$.  
The probability that the atom survives this truncation process
is given by Eq.~(\ref{eqn:ProbSur}) as $P_{\rm surv}(U_{\rm trunc})$. 
Naturally, this probability drops quickly as $\eta \to 0$, thus reducing 
the efficiency (or effective duty cycle) of this scheme. 

Figure~\ref{fig:PostSelectResults}(d) shows the mean energy calculated using
Eq.~(\ref{eqn:MeanEnergy}) with the measured values of the temperature and 
the truncation energy $U_{\rm trunc}$. 
The solid line is a theoretical prediction using Eq.~(\ref{eqn:MeanEnergy}) for a 
temperature $T =33~\mu$K.
 
For minimum trap depths above $10~\mu$K, i.e. when the energy distribution is 
approximately non-truncated, the mean energy is $\langle E\rangle\sim100~\mu{\rm K}$. 
For minimum trap depths below $10~\mu$K, we see that the mean energy is significantly 
reduced. For example, for a minimum trap 
depth $U_{\rm min}\sim 0.4~\mu$K, we measure a mean energy  of the atom 
at the end of this filtering sequence of $\langle E\rangle\sim30~\mu{\rm K}$, 
where the final trap depth is $U_{\rm f}\sim2.8$ mK.  
The truncation of the Boltzmann distribution thus leads to a reduction of the mean 
energy by a factor of $100\mu{\rm K}~/~30\mu{\rm K}\sim3.3$. 
At the same time, the probability of recapture is reduced to $P_{\rm rr}(\Delta{t}=0)\sim0.1$. 

In this section we have shown that if one is willing to compromise on the probability 
of recapturing the atom (or correspondingly, the duty cycle of the experiment), 
truncating the Boltzmann distribution is a good way to reduce the mean energy 
of the atom. 

\section{Cooling a Single Atom by Adiabatically Lowering the Trapping Potential} \label{sec:AdiabCool}

Let us consider an atom in a trapping potential. By adiabatically lowering 
the trap depth the occupation probabilities of the vibrational levels are 
preserved \cite{Reif}. For our optical dipole trap this conservation of the vibrational 
number results in $T/\sqrt{U}$ being a constant. 
In this section we will experimentally demonstrate that by slowly lowering the 
trap depth, the temperature of the atoms follows this relation, 
thus reinforcing our assumptions about adiabaticity. 

\begin{figure}
\begin{center}
\includegraphics[width=8.5cm]{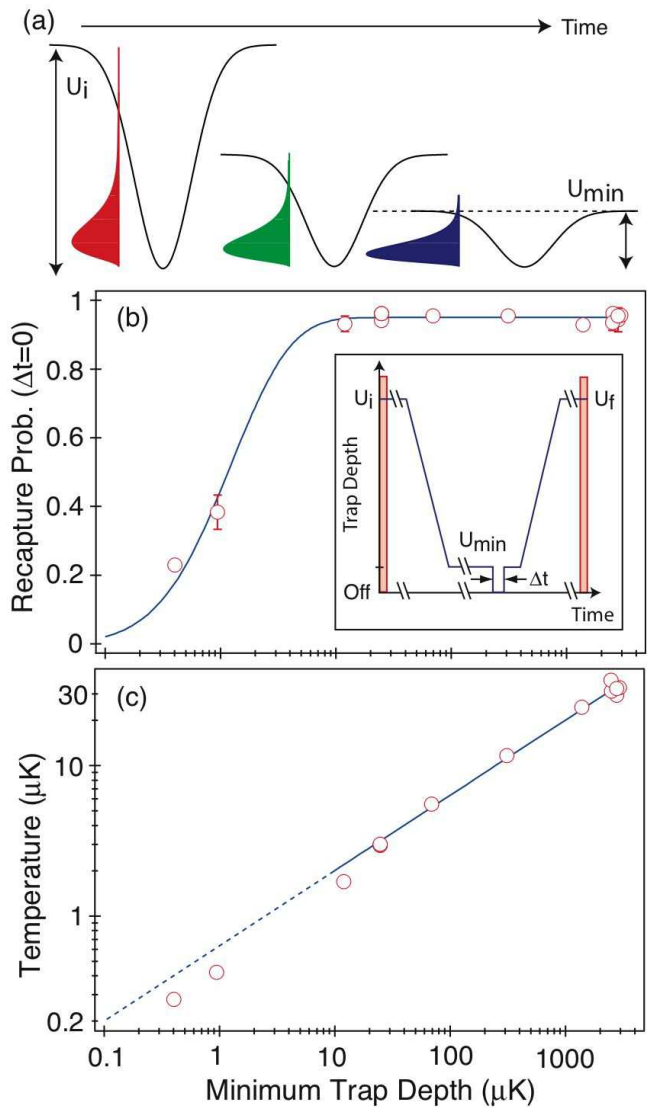}
\caption{(Color online) Investigation of the adiabaticity during the change of the optical trap depth. 
(a) Schematic of the experimental sequence (see text). 
(b) Probability of recapture for zero release time $P_{\rm rr}(\Delta{t}=0)$, 
as a function of the minimum trap depth $U_{\rm min}$. 
The solid line is a theoretical fit to the data.
Inset shows a schematic of the temporal experimental sequence. 
(c) Temperature $T$ of the Boltzmann distribution plotted
as a function of $U_{\rm min}$. The solid line is a theoretical 
prediction of the adiabatic scaling law $T/\sqrt{U} =$~const.
 The dashed line is the extension of the solid line to minimum 
trap depths at which the density of states deviates from 
that of a three dimensional harmonic trap.
} \label{fig:AdiabaticResults}
\end{center}
\end{figure}

To test the adiabaticity of our lowering scheme we perform the same experimental 
sequence as in section~\ref{sec:TrucBoltz}, but we now implement the ``release and 
recapture" sequence when the trap is shallowest, i.e. we measure the temperature 
of the atoms at the minimum trap depth $U_{\rm min}$~\footnote{After 
the ``release and recapture'' sequence we 
needed to bring the trap depth close to its initial value to 
measure if the atom is still present in the trap, otherwise the lasers sent on the 
atom would kick it out of the shallow trap too quickly to measure its presence.}. 
This is in contrast to section~\ref{sec:TrucBoltz}, where the temperature of the 
atoms was measured after the 
trap depth was returned to its initial value. 

A schematic of the experimental sequence is shown in Figure~\ref{fig:AdiabaticResults}, 
together with the recapture probability, $P_{\rm rr}(\Delta{t}=0)$, and the temperature, $T$, measured 
for various minimum trap depths $U_{\rm min}$. The data in figure~\ref{fig:AdiabaticResults}(c), 
shows evidence that the temperature of the Boltzmann distribution follows closely 
an adiabatic lowering behavior. For comparison, we have plotted
the adiabatic scaling law $T/\sqrt{U}=$~const, where the constant is equal to  
$T_{\rm i}/\sqrt{U}_{\rm i}$ with $T_{\rm i}=33~\mu{\rm K}$ and $U_{\rm i}= 2.8$ mK. 

The criteria for adiabaticity states that the  rate of change of the oscillation 
frequency $\dot{\omega}$ must be lower than $\omega^2$ ($\omega$ is the 
oscillation frequency) at all times during the evolution~\cite{Landau}. 
Care must be taken to fulfill the adiabaticity criteria when the trapping 
potential is very small, corresponding to a small oscillation frequency. The rate of 
change of the oscillation frequency must be sufficiently small in this case.
In our experiment, this is done by smoothly ramping up and down 
the intensity of the dipole trap laser with acousto-optic modulators. The behavior
of the measured temperature in Figure~\ref{fig:AdiabaticResults}(c) 
gives us a strong indication that indeed the adiabaticity criteria is 
fulfilled when the trapping depth is small.

Figure~\ref{fig:AdiabaticResults}(b) shows the probability of recapture 
for zero release time $P_{\rm rr}(\Delta{t}=0)$ plotted as a 
function of the minimum trap depth $U_{\rm min}$, together with a theoretical 
fit.  We reemphasize that for minimum trap depths less than $U_{\rm min}\sim10~\mu{\rm K}$, 
the Boltzmann distribution becomes truncated. This implies that 
we are no longer in the three-dimensional harmonic approximation, and hence 
the density of states is no longer described by 
$\rho(E)\propto{E}^2$.
Nevertheless, for clarity in Figure~\ref{fig:AdiabaticResults}, we have presented 
the results assuming that $\rho(E)\propto{E}^2$ for all data points. 

As a typical result, we measure a temperature of $T=1.69\pm{0.02}~\mu{\rm K}$
for the atoms in a trap that has been adiabatically lowered to a depth of 
$U_{\rm min}\sim12~\mu{\rm K}$. 
In this case, the Boltzmann distribution is only slightly truncated, implying that 
the quadratic density of states should still be valid. This temperature corresponds 
to a one hundred fold decrease in the 
temperature of the atoms compared to those that 
were directly loaded from the molasses. 

\section{Discussion} \label{Discussion}

We have shown that the single atom cooling via filtering techniques is quite efficient. 
Starting from an initial mean energy  of 100 $\mu$K after the laser cooling phase, 
we get a final mean energy of the atom of 30~$\mu$K 
with a probability of $\sim0.1$. On our experiment, the loading rate of single atoms 
is on the order of 1 per second. If we are willing to comprise our duty cycle, 
we could obtain on average one atom every 10 seconds
with a mean energy of 30~$\mu$K in a $\sim2.8$~mK deep trap, which corresponds 
to a ratio of the trap depth to mean energy 
of $\sim93$. 

Another parameter characterizing the external motion of the atom in the 
trap is the mean vibrational number along one axis,
defined as the ratio of the mean energy of the atom along this axis
to the energy difference between vibrational states. As the trap has two 
different oscillation frequencies along the axial and radial
directions, this corresponds to  two different mean vibrational numbers along these directions. 
If we assume the equipartition of the mean energy among the three axis, we get for the mean vibrational
number along  the radial direction $n_{\perp}= \langle E\rangle/ 3 h\nu_{\perp}$. 
For the mean energy of $100~\mu$K that we obtained after laser-cooling the atom, this 
leads to $n_{\perp}\sim4.3$. 
By contrast, the lowest mean energy of  $30~\mu$K that we obtained after 
filtering leads to $n_{\perp}\sim1.3$, which 
is $3.3$ times smaller than the mean vibrational number before filtering. 
This mean number, close to 1 along the radial direction, raises some 
questions about the validity of our classical treatment for the temperature 
measurement for the lowest trap depths. This analysis, albeit interesting, 
is beyond the scope of this paper. This also places us in a good position 
to further cool the atom down to the ground state, for example by using Raman 
sideband cooling~\cite{Diedrich89}.

In the framework of quantum information, there has been proposed a 
protocol to entangle two atoms 
based on the emission of a single photon by one of the two trapped atoms~\cite{cabrillo99}.
A crucial feature of  this proposal is the ability to localize each atom 
within a distance smaller than the wavelength of the emitted photon, 
the so-called Lamb-Dicke regime.  Reference~\cite{cabrillo99} shows 
that for atoms described by a thermal distribution at temperature $T$, 
this condition becomes $\eta_{\rm th}~=~\eta_{\rm LD} \sqrt{k_{\rm B}T/\hbar\omega}~\ll~1$, 
with $\omega$ being the mean oscillation frequency of the atom in the trap, 
and $\eta_{\rm LD}~=~\sqrt{E_{\rm r}/\hbar\omega}$ being the Lamb-Dicke parameter ($E_{\rm r}$ is the
recoil energy). In our case, for a trap depth of $\sim2.8$~mK, we find $\eta_{\rm LD}~=~0.20$  and $\eta_{\rm th}~=~0.56$ for a temperature of 
33 $\mu$K, as measured above. We are therefore in the right regime to 
apply this entanglement protocol. We finally note that $\eta_{\rm th}$ can be lowered 
when increasing the trap depth adiabatically. In this case $\eta_{\rm th}$ is related to the 
trap depth $U$ by  $\eta_{\rm th}\propto U^{-1/4}$.

\section{Conclusion}\label{Conclusion}

As a summary of this paper, we have investigated the energy of a 
single atom trapped in a tight optical dipole trap. We have found that 
after applying  laser cooling of the atom in the tweezer, the energy  
of the trapped atom follows  a Boltzmann distribution.    We have extracted 
the temperature of the atoms from the measured energy distribution.
We have also described a  ``release and recapture'' method to measure the 
temperature of the atoms assuming a truncated Boltzmann distribution. 
This second method is in good agreement with the first one.

We have shown that by adiabatically lowering the trap depth we 
have reduced the temperature of the atoms down to several microKelvin. 
Finally, we have cooled the atom by filtering the hottest atoms over many 
realizations of the same experiment, the equivalent of evaporative cooling 
but at the single atom level. This leads to an atom close to the ground state in the 
radial direction of the trap.

\begin{acknowledgments}
We acknowledge support from the Integrated Project `SCALA' which is part of 
the European IST/FET/QIPC Program, from Institut Francilien des Atomes
Froids (IFRAF), from ARDA/DTO, and from PPF-ENS. Y.~R.~P.~Sortais 
was supported by IFRAF and CNRS fellowship, and A.~M.~Lance also by IFRAF and by a Marie Curie fellowship.
\end{acknowledgments}

\end{document}